\documentclass[
  journal=ram,
  manuscript=article-type,
  year=,
  volume=,
]{cup-journal}

\usepackage{amsmath}
\usepackage[nopatch]{microtype}
\usepackage{multirow}
\usepackage{booktabs}
\usepackage[hidelinks, breaklinks=true]{hyperref}
\usepackage{setspace}
\title{MDAF: A Multi-Dimensional Annotation Framework for Automated Foreign Policy Analysis}

\author{Songruowen Ma}
\affiliation{Department of International Development, University of Oxford, Oxford, OX1 3TB, United Kingdom}

\author{Songting Ding}
\affiliation{Graduate School of Frontier Sciences, University of Tokyo , Tokyo, 2770882, Chiba, Japan}
\email[Corresponding Author]{songting.ding@gmail.com}

\author{Fangyin Zhou}
\affiliation{School of International Relations, Sun Yat-sen University, Zhuhai, 519082, Guangdong, China}

\author{Jörg Friedrichs}
\affiliation{Department of International Development, University of Oxford, Oxford, OX1 3TB, United Kingdom}


\usepackage{booktabs}
\usepackage{longtable}
\usepackage{array}
\usepackage{ragged2e}

\begin{document}
\begin{abstract}
Government websites contain a vast but underexploited body of textual evidence on foreign policy. This article develops a scalable approach for extracting structured information from policy texts and converting it into standardized event data, with policy event defined broadly as a statement or action. It proposes MDAF integrating an LLM workflow to automate foreign policy text identification, information extraction, and event classification. Empirically, it applies this approach to China-related texts from Five Eyes countries. The analysis shows that the resulting database supports systematic cross-national comparison, reveals variation in how states frame and implement China policy, and traces the temporal evolution of these policy profiles. This article contributes to foreign policy analysis and the methodological development of computational international relations.
\end{abstract}
\section{Introduction}
Government websites constitute an important public-facing medium through which states articulate policy positions, define external relationships, and communicate strategic intent. Texts published on these websites therefore offer a valuable empirical basis for identifying policy stances and tracing their evolution in international relations research. Whether scholars examine interstate cooperation and conflict, diplomatic narratives, or policy change, they often rely on such texts to trace how states respond and how their behavior evolves over time \autocite{smith2021MS,blomquist2026MS,deyermond2023MS}.

However, policy texts are rarely reducible to a single position or event (which we define broadly as a policy statement or policy action). A single document may combine normative statements, concrete policy measures, relational framings, and signals of cooperation, competition, or confrontation.\footnote{Different orientations may coexist within the same text. For instance, in a 2024 meeting with China’s Foreign Minister, the UK Foreign Secretary stated that the government would “co-operate where it can, compete where it must, and challenge where it must.” He welcomed opportunities for cooperation with China on issues such as climate change, while also urging China to prevent its companies from supporting Russia’s military-industrial complex, which he described as posing a material threat to international security and prosperity. See UK Government, “Readout of Foreign Secretary meeting with China’s Director of Foreign Affairs Commission Office and Foreign Minister,” 26 July 2024, \href{https://www.gov.uk/government/news/readout-of-foreign-secretary-meeting-with-chinas-director-of-foreign-affairs-commission-office-and-foreign-minister}{https://www.gov.uk/government/news/readout-of-foreign-secretary-meeting-with-chinas-director-of-foreign-affairs-commission-office-and-foreign-minister} } While manual coding remains viable for small-scale research, another methodological challenge is how to preserve the richness of official policy texts while transforming continuously updated and structurally diverse documents into standardized, comparable, and scalable data. This challenge bears directly on measurement quality, cross-national comparison, and cumulative knowledge production in foreign policy analysis.

Existing research has made important contributions to the structured recording and automated identification of political events, state behavior, and policy texts, providing a foundation for this study. But they remain limited when applied to official policy texts that are continuously updated, semantically complex, and strategically oriented.

First, some influential datasets still rely heavily on manual coding, which constrains large-scale and high-frequency updating. For example, the Militarized Interstate Disputes (MIDs) dataset and the Manifesto Project mostly depend on trained researchers to verify and code information such as actors, time, event types, and outcomes \autocite{werner2021MS}. This approach improves accuracy and interpretability, but it also entails high production costs and long update cycles.

Second, many existing datasets are designed for specific research purposes. They select and standardize the elements most relevant to a particular empirical phenomenon, while leaving other information contained in policy texts outside the scope of data collection. For example, while conflict and political violence datasets such as Armed Conflict Location and Event Data Project (ACLED) and Uppsala Conflict Data Program (UCDP) are highly valuable for tracing conflict processes, they generally do not incorporate dimensions related to policy motivation. Consequently, they are unable to provide adequate empirical support for research on issues such as the justificatory logic of state behavior. This limitation reduces the usefulness of such datasets for studies seeking a more comprehensive representation of foreign policy stances across issue areas, cases, and time \autocite{acled2024MS,hogbladh2025MS}.

Third, existing event data systems remain only partially connected to core theoretical concepts in international relations. Systems such as the Integrated Crisis Early Warning System (ICEWS) and the Global Database of Events, Language, and Tone (GDELT) organize textual information around event records, typically specifying who did what to whom\autocite{lockheedmartinICEWS, gdelt2015Codebook, schrodt2012CAMEO}. This event-centered logic is valuable for standardizing political interactions, but less suited to capturing how textual and behavioral indicators map onto concepts such as state intention, threat perception, identity recognition, partner-competitor positioning, and policy adjustment. As policy texts continue to grow in volume and complexity, researchers need not only more efficient automated tools, but also a coding framework that links textual attributes, behavioral identification, and theoretically meaningful concepts.

\begin{figure}[htbp]
    \centering
    \includegraphics[width=1\textwidth]{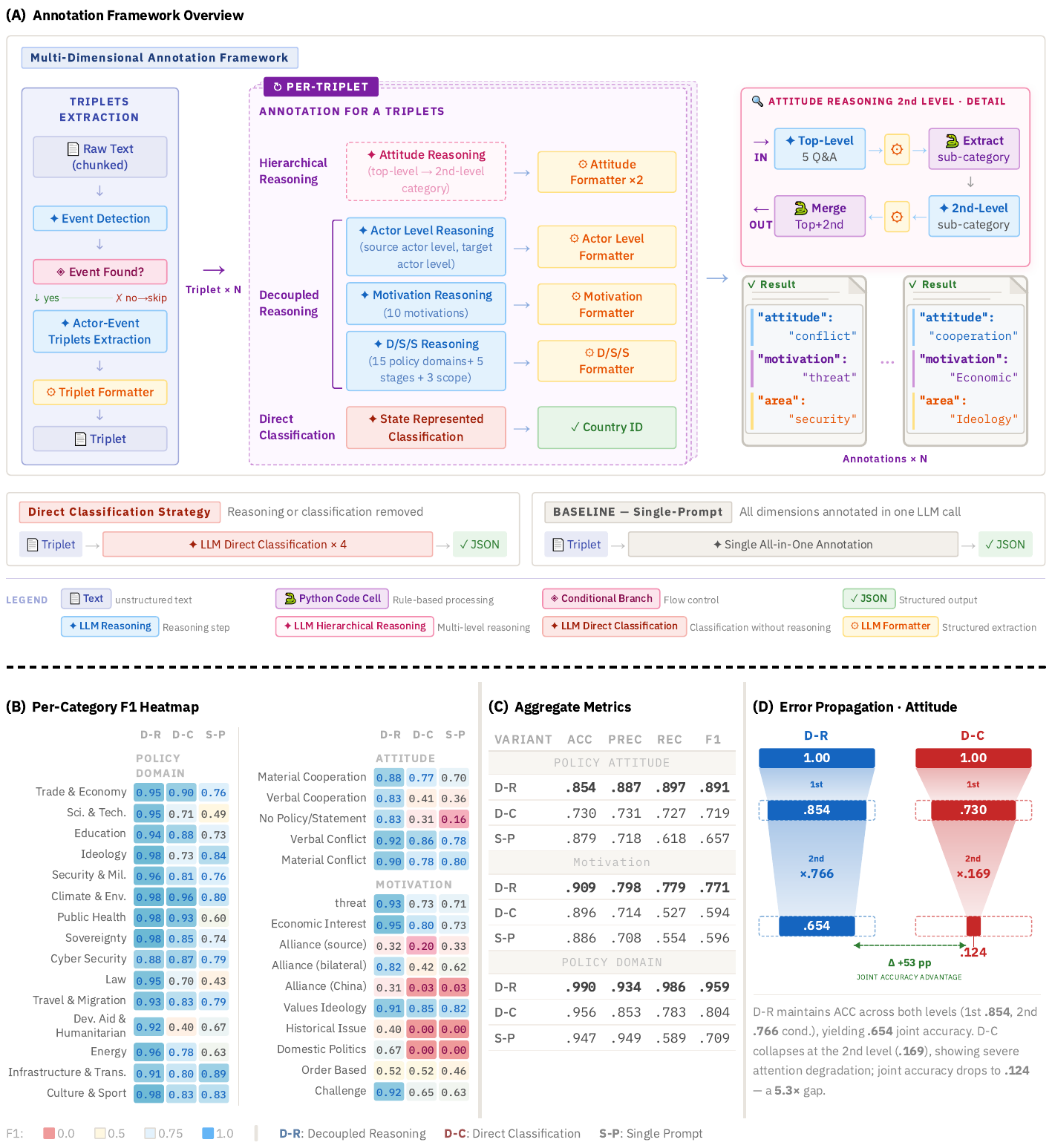}
    \caption{Multi-Dimensional Annotation Pipeline and Performance Evaluation
    \textit{(A) The proposed pipeline decomposes the annotation process into three strategies, providing a structured and decomposed reasoning framework. To demonstrate the effectiveness of MDAF, we compare it with two alternative variants: a direct-classification ablated variant and a single-prompt baseline variant. (B-C) Per-category and aggregate F1 comparisons show that decoupled reasoning achieves consistently better performance than the alternative strategies across the attitude, motivation, and policy domain dimensions. (D) Error propagation analysis shows that decoupled reasoning experiences a sharp decline in performance relative to hierarchical reasoning, highlighting the value of hierarchical reasoning in large-label-space classification tasks.}}
    \label{fig.1}
\end{figure}

To address such limitations, this article introduce an annotation framework that takes the actor-event as the core unit of analysis. For each foreign policy text, the framework records features across multiple dimensions, including the actor, policy domain, scope, stage, attitude, and motivation. This design allows us to characterize the content and patterns of foreign policy in a more systematic way. To implement this framework, we introduce a reasoning-augmented LLM annotation workflow. The workflow decomposes the annotation process into a sequence of linked tasks, including document identification, information extraction, and event classification. This task decomposition improves the accuracy, scalability, and reproducibility of automated coding in complex policy contexts. Figure \ref{fig.1}, which will be fully explained in the course of this article, provides an initial overview of the annotation pipeline and summarizes the comparative performance of the reasoning-augmented approach against alternative prompting strategies.

The article makes three contributions. Methodologically, it proposes an annotation framework designed to make fuller use of the rich information embedded in official policy texts and to support fine-grained comparison across different analytical dimensions. Technically, it demonstrates how large language models, when combined with structured reasoning prompts and task decomposition, can be used for automated coding of complex foreign policy documents. Empirically, it shows how the proposed workflow can generate structured data for comparing national policy profiles and tracing their evolution over time. Our aim is not simply to introduce a coding scheme or an annotation procedure. Rather, we seek to build an end-to-end system for policy quantification, from text collection to label generation: modular, extensible, and capable of evolving with new research questions.

\section{Multidimensional Annotation Based on Actor-Event Triplets}
We adopt a two-step process for policy document coding. In the first step, we identify and extract actor-event triplets from original texts. An actor-event triplet refers to a specific policy event, namely a policy statement or policy action, issued by a source actor and referring to a target actor within a given textual segment, or chunk. In the second step, we treat each actor-event triplet as an annotation unit and code it through a multidimensional policy coding framework. In this way, unstructured text is transformed into structured data that can be compared and aggregated. The framework further identifies for each actor-event triplet the source actor’s attitudes and motivations as well as the relevant policy domain, policy stage, and policy scope.

\subsection{Extracting Actor-Event Triplets}
In the tradition of political event data research, event records are fundamentally structured around the question of “who did what to whom”. From Conflict and Mediation Event Observations (CAMEO) to Political Language Ontology for Verifiable Event Records (PLOVER), event coding frameworks have remained centered on the “source actor-event-target actor” triplet as the basic unit of analysis \autocite{gerner2002CAMEO,halterman2023MS}.

The multidimensional policy annotation framework inherits this logic: a concrete policy action initiated by a specific actor and directed towards another actor serves as the foundational carrier for subsequent annotation dimensions, including attitudes, motivations, and policy domains. In other words, attitudes and motivations are not attached to an entire document or text segment, but rather to a specific policy action performed by a clearly identifiable actor towards a clearly identifiable target.

Within our framework, the annotation unit is defined as an actor-event triplet. However, accurately extracting structured information from policy documents remains highly challenging. Policy texts frequently contain multiple actors and multiple policy events intertwined within the same paragraph or even the same sentence. Conventional political event coding systems, such as Textual Analysis By Augmented Replacement Instructions (TABARI), primarily rely on syntactic parsing \autocite{schrodt2009TABARI}.  Although effective for relatively simple event structures, these approaches suffer substantial recall degradation when multiple events coexist within long and semantically dense documents. Recent neural and LLMs-based extraction methods demonstrate stronger semantic comprehension capabilities, yet multi-event contexts still introduce substantial risks of cross-event argument confusion, where actors belonging to one event may be incorrectly associated with another event.

To address this challenge, we design event detection as an independent preprocessing stage. The purpose is to identify and isolate each policy event before multidimensional annotation begins. Instead of directly annotating a text chunk containing multiple overlapping events, the system first detects each individual event and explicitly determines its source actor and target actor. By decomposing complex text chunks into independent actor-event pairs, the framework substantially reduces the risk of ambiguity in actor attribution under multi-event conditions.

\subsection{Annotating Actor-Event Triplets Across Seven Dimensions}
This section introduces the annotation scheme used to classify the extracted actor-event pairs. The scheme includes seven dimensions. Five focus on the policy: domain, attitude, motivation, stage, and scope. These dimensions capture what the action concerns, how it is oriented, how it is justified, how far it has advanced, and what interactional setting it refers to. The remaining two dimensions, actor level and state represented, describe the actors associated with the event, identifying their institutional position and national affiliation. Online Appendix A reports the full label definitions and annotation rules.

\subsubsection{Policy Domain}
Policy domain identifies the issue area a policy statement or action refers to. Foreign policy is not a single, homogeneous field. Different domains involve different instruments, interaction patterns, and theoretical logics. Security and military issues, for example, are often connected to threat perception, deterrence, and alliance politics, whereas trade, economic, and investment issues are more closely associated with interdependence, market access, and industrial competition. Climate change, public health, and development assistance often reflect transnational governance and collective-action problems. Annotating policy domain makes it possible to capture issue-level variation within official texts and convert that variation into an object of comparative analysis.

Issue classification is a core staple in the study of comparative political texts. For example, the Manifesto Project codes party manifestos at the level of quasi-sentences and classifies them across seven policy domains: external relations, freedom and democracy, political system, economy, welfare and quality of life, fabric of society, and social groups \autocite{werner2011MS}. The Comparative Agendas Project similarly classifies policy-related observations into 21 major topics and 220 subtopics \autocite{capMS}.

Building on existing practices of issue classification and the policy categories commonly used in official government sources, this article identifies the following policy domains: climate change and environment; culture, media, and sport; cybersecurity; development assistance and humanitarian aid; education; energy; ideology; infrastructure and transportation; law; public health; science, technology, and innovation; security and military affairs; sovereignty; trade, economy, and investment; travel; immigration and refugees.

\subsubsection{Policy Attitude}
Policy attitude is central to the analysis of state behavior because it captures how a source actor positions itself towards a target actor. Without further differentiation, researchers may misread the direction of behavior, misjudge the intensity of change, or conflate symbolic rhetoric with concrete action.

International relations event data have long relied on cooperation-conflict scales as a measurement strategy. The World Events Interaction Survey (WEIS) project arranges diplomatic interactions across 22 categories ranging from cooperation to conflict \autocite{goldstein1992MS}.  The Conflict and Peace Data Bank (COPDAB) assigns numerical scale values to international events along a conflict–cooperation continuum \autocite{azar1980MS}.  GDELT has incorporated a similar measurement strategy through the Goldstein Scale, which assigns each event type a score from -10 to +10 to indicate its expected effect on the stability of the target country, with negative values denoting more conflictual events and positive values denoting more cooperative ones \autocite{gdelt2015Codebook}. 

Building on this tradition, we classify policy attitude into five first-level categories: material cooperation, verbal cooperation, no policy or statement, verbal conflict, and material conflict. Material cooperation captures cooperative measures that have been implemented or are being advanced, including the establishment or upgrading of bilateral relations, the signing of agreements, expanded exchanges, friendly military activities, or the removal of restrictions. Verbal cooperation refers to cooperative positions expressed through approval, welcome, appreciation, or support, in the absence of corresponding concrete policy action. No policy or statement covers texts that either express a position without a clear cooperative or confrontational orientation, or provide only informational or procedural notice without obvious policy significance. Verbal conflict captures negative positions conveyed through concern, regret, dissatisfaction, opposition, condemnation, warning, or protest, without accompanying confrontational measures. Material conflict refers to confrontational policy measures that have been adopted or are being advanced, such as downgrading relations, restricting exchanges, weakening existing agreements, undertaking hostile military actions, or intensifying restrictive measures. Each first-level category is further specified through a set of second-level subcategories, reported in Online Appendix A, to capture finer distinctions in policy expression.

\subsubsection{Policy Motivation}
Policy motivation captures the rationale a text attaches to a policy position or action. Official policy documents rarely state only what a government does; they often also explain why a particular position is taken, whether in terms of security concerns, normative commitments, alliance obligations, domestic considerations, or claims about international order. Coding these justifications matters because similar policy actions may carry different meanings depending on the considerations invoked to support them.

Most existing event-data systems do not capture this layer of meaning directly. For instance, CAMEO-based systems, including GDELT, classify political events and their general conflictual or cooperative implications, while GDELT’s Goldstein Scale and tone measures capture event-level impact or textual sentiment rather than the reasons invoked for a policy action. This article annotates policy motivation separately, focusing on the justificatory claims through which governments explain their policy positions.

The motivation categories are derived from major theoretical traditions in international relations and foreign policy analysis. Realist and alliance-based approaches emphasize security threats, power competition, and alliance constraints; liberal and institutionalist perspectives highlight economic interdependence, shared challenges, and institutional cooperation; constructivist approaches foreground identity, values, historical memory, and claims about international order; and foreign policy analysis draws attention to domestic politics, leadership legitimacy, and internal agenda-setting. These traditions provide the theoretical basis for distinguishing the different justificatory logics through which governments explain their policy positions.

On this basis, we classify policy motivation into nine categories: alliance-related, shared-challenge, domestic-political, economic-interest, historical-issue, order-related, threat-based, value- and ideology-based, and other. Online Appendix A provides detailed descriptions.

\subsubsection{Policy Scope}
Policy scope identifies the institutional setting in which a policy position or action is articulated. The same policy content may carry different constraints, sources of legitimacy, and political meanings depending on whether it is advanced unilaterally, negotiated bilaterally, or embedded in a multilateral framework. Sanctions may be imposed by a single state or coordinated through a multilateral mechanism; a cooperative initiative may target one partner or operate through a regional or global governance arrangement. Major event-data projects, including Political Event Classification, Attributes, and Types (POLECAT), GDELT, and ICEWS, are designed to capture policy events without coding scope as a distinct analytical dimension.

We classify policy scope into three categories: unilateral, bilateral, and multilateral. Unilateral policies are undertaken independently, without joint coordination or formal interactive arrangements. Bilateral policies are typically conducted between the two actors of an actor-event triplet. Multilateral policies are advanced through international organizations, regional mechanisms, or other frameworks involving multiple actors. This distinction helps identify the institutional setting in which policy action is articulated and the strategic meanings attached to different forms of engagement.

\subsubsection{Policy Stage}
Policy stage captures the degree to which a policy action has moved from expression to execution. Official policy texts do not always describe fully implemented measures. Some state only principles or intentions, while others announce preparatory work, specify forthcoming action, report ongoing implementation, or confirm completed outcomes. Event-data systems such as GDELT are useful for identifying types of political action and their conflictual or cooperative implications, but they do not record how far a policy commitment has advanced. Without such a distinction at the annotation level, researchers may overestimate the effect of policy commitments and treat symbolic statements as equivalent to implemented policies.

We classify policy stage into five categories: pure intention, preliminary planning, imminent action, ongoing, and completed. Pure intention refers to statements of principle or directional expressions. Preliminary planning captures initial design, discussion, or preparation. Imminent action refers to cases in which the action has a clear expected timeline or implementation arrangement. Ongoing refers to policies that are being advanced or implemented. Completed refers to measures that have already been carried out or outcomes that have been confirmed. This distinction allows us to record not only the substance of a policy, but also its level of advancement.

\subsubsection{Actor Level}
Actor level identifies the institutional position of the actor in a policy text. The meaning of an official statement depends not only on what is said, but also on the authority and representational status of the actor that issues it. Heads of state and governments often speak closer to the level of national strategy; ministries tend to articulate sector-specific positions; legislatures may reveal domestic political constraints; and subnational or non-state actors usually carry a different degree of national representativeness. Event datasets like GDELT and ICEWS identify the actors involved, but they are not designed to annotate the institutional level from which a policy signal is issued. Without this distinction, researchers may treat highly heterogeneous statements as equivalent observations and misread their authority, representativeness, or relevance to national policy.

We classify actor level into seven categories: head of state; national state; national ministry; legislative body; subnational government; intergovernmental organization or forum; and non-state actor. This dimension captures the institutional location of policy signals and supports the analysis of signal strength, political source, and national representativeness.

\subsubsection{State Represented}
This dimension identifies the country or political entity associated with the actor in a policy text. Official texts do not always refer to actors through standardized country names. Instead, they may mention named leaders, presidents, prime ministers, foreign ministries, defense ministries, parliaments, embassies, government delegations, or other institutions. Without mapping these dispersed references to a common national unit, it is difficult to determine whose policy position a statement represents or to compare policy signals systematically across countries, time periods, and issue areas.

This dimension follows the standardization logic used in event-data research. Datasets such as ICEWS and GDELT map specific textual entities onto country codes, organizational categories, or political entities to produce comparable units of observation. Building on this approach, we link concrete offices, institutions, and delegations in official texts to their corresponding state or political actor. For example, the U.S. president, the U.S. Department of State, U.S. embassies abroad, and U.S. delegations are all coded under “United States”. This procedure preserves information about the specific actor while generating a stable state-level identifier for country-level aggregation, time-series analysis, and cross-national comparison.

\section{Structured and Decomposed Reasoning Framework}
Chain-of-Thought (CoT) is a prompt engineering strategy that improves the performance of large language models on complex tasks by eliciting intermediate reasoning steps. \textcite{wei2022MS} were the first to systematically demonstrate that incorporating step-by-step reasoning into prompts can significantly boost LLM accuracy on arithmetic, commonsense reasoning, and symbolic reasoning tasks.  In the domain of political text analysis, recent work has begun to explore the potential of CoT, suggesting that CoT reasoning holds promise for improving over zero-shot LLM performance on complex codebooks \autocite{halterman2026MS}. \textcite{tornberg2025MS} confirms that LLMs augmented with step-by-step reasoning can surpass expert coders when annotating political social media texts. 

The underlying problem is that multi-dimensional annotation imposes substantially greater reasoning burdens on LLMs than conventional single-task classification settings. The difficulty stems primarily from two aspects:

The first problem is that a single policy event simultaneously encompasses multiple dimensions, such as entities, attitudes, motivations, and policy domains, which are logically interrelated (e.g., certain attitudes tend to correspond to certain motivations). Requiring the model to output all labels at once readily leads to cross-dimensional interference and omission.

The second problem is that the sub-tasks involved in multi-dimensional policy annotation vary substantially in cognitive complexity. Actor recognition typically relies on locally explicit information, whereas identifying behavioral motivations often requires cross-sentence semantic integration and implicit intent inference. A uniform CoT pipeline is therefore ill-suited to accommodate all dimensions simultaneously. What is needed is a structured framework capable of dynamically adjusting reasoning paths according to task complexity, so that different CoT procedures can be appropriately allocated.

\begin{figure}[htbp]
    \centering
    \includegraphics[width=1\textwidth]{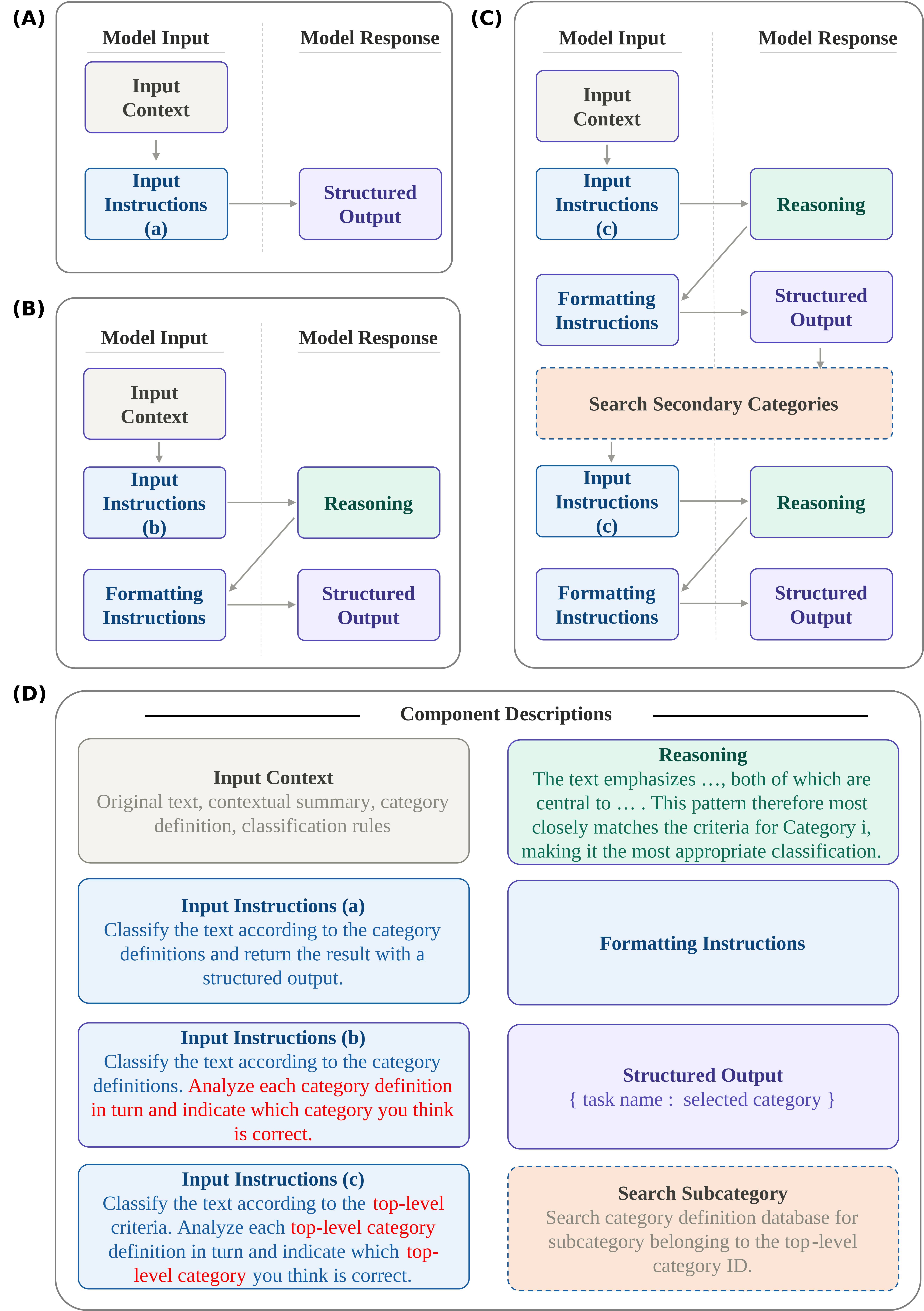}
    \caption{Structured and Decomposed Reasoning Framework}
    \label{fig.2}
\end{figure}

As illustrated in Figure \ref{fig.2}, we propose Multi-Dimensional Annotation Framework (MDAF) as a structured and decomposed reasoning tool for policy annotation using LLMs. We conceptualize multi-dimensional policy annotation as a heterogeneous reasoning problem, where different annotation tasks impose fundamentally different inference burdens on LLMs.

The core design principle of the framework is structured decomposition. Rather than treating policy annotation as a single end-to-end classification task, the framework decomposes the process into different reasoning strategies. Specifically, MDAF encompasses direct classification, decoupled reasoning, and hierarchical reasoning to accommodate challenges of varying difficulty. These challenges necessitate a structured reasoning framework that allocates different reasoning strategies according to the inference burden of each annotation task.

\subsection{Direct Classification for Low-Inference-Burden Tasks}
For basic information extraction tasks LLMs already achieve very high accuracy. \textcite{lee2025GPT} find in political science research that ChatGPT-4 exhibits high accuracy when extracting descriptive information such as time and location, with error rates below 5\%, meaning that additional reasoning steps yield only marginal gains. 

In our multi-dimensional framework, state represented is precisely this kind of low-inference-burden task. State represented requires no deep semantic analysis and can be extracted directly from the text based on its literal meaning. This means the mapping can be completed through direct output, without the need for any additional reasoning-formatting decoupling. Based on this judgment, we adopt direct annotation for the state represented dimension.

The pipeline for the direct classification strategy is shown in Figure 2(A). In MDAF, the data inputted into the framework are carefully structured. We provide the original text of actor-event triplet. To avoid the loss of contextual information, such as event background, referential links, irony, or other context-dependent meanings, we also use an LLM to generate a very brief summary of the text before or after the input triplet. This summary, together with the original text, detailed definitions, and supplementary rules for each category, is provided to the LLM as the input context. The LLM is then prompted to directly predict the category ID (for the representative state dimension, this refers to the country or political entity name) based on the input context.

\subsection{Decoupled Reasoning for Medium-Inference-Burden Tasks}
For medium-inference-burden tasks, correct classification depends on semantic reasoning rather than simple pattern matching of text. At this level, tasks move beyond straightforward information extraction and require the model to distinguish among multiple semantically related categories. When an LLM is asked to simultaneously perform semantic reasoning and produce structured output, the two demands interfere with each other: the attentional resources allocated to following format constraints crowd out those available for semantic judgment \autocite{halterman2026MS,tam2024Let}. 

In our framework, policy domain, motivation, stage, scope, and actor level fall into this category of medium-inference-burden tasks. These dimensions share a common profile: a moderate number of candidate categories (5-15), partially overlapping category boundaries, and a requirement for deep semantic understanding or at least rudimentary reasoning to arrive at the correct answer. 

We apply the decoupled reasoning strategy uniformly to all such tasks, separating the semantic reasoning step from the procedural formatting step. As illustrated in Figure 2(B), each annotation task is split into two separate LLM calls. The reasoning step first asks the model only to answer a prompt in natural language and articulate the basis for its judgment, imposing no output format constraints whatsoever. The formatting step, then, takes the reasoning output as its input and converts it into structured output. This separation yields two benefits: the quality of the reasoning-stage output is no longer compromised by formatting constraints, while the input for the formatting step is already prestructured text that has been reasoned through, substantially reducing the difficulty of extraction. We adopt the decoupled reasoning strategy as the general-purpose approach for all medium-inference-burden tasks.

\subsection{Hierarchical Reasoning for Large-Label-Space Tasks}
Hierarchical reasoning has been a long-standing research challenge in machine learning. In the computer vision literature, researchers have long demonstrated that a strategy of first classifying at a coarse-grained level and then refining to sub-categories outperforms direct prediction of leaf nodes \autocite{deng2014MS}.  In text classification, hierarchical multi-label classification is likewise an active area of inquiry - when a label taxonomy contains hundreds or even thousands of categories, including the full set of label definitions in the prompt rapidly consumes tens of thousands of tokens of context window, leading to significant degradation in LLM performance \autocite{tabatabaei2025MS}.  \textcite{halterman2026MS}, in experiments on a party manifesto dataset containing 142 categories, found that even state-of-the-art LLMs perform extremely poorly when confronted with lengthy codebooks, with the likely explanation being attentional decay towards content in the middle of the prompt.  This finding suggests that once a classification taxonomy exceeds a certain scale, a “flat” direct classification strategy is no longer viable, and hierarchically structured step-by-step reasoning becomes necessary.

\begin{figure}[htbp]
    \centering
    \includegraphics[width=1\textwidth]{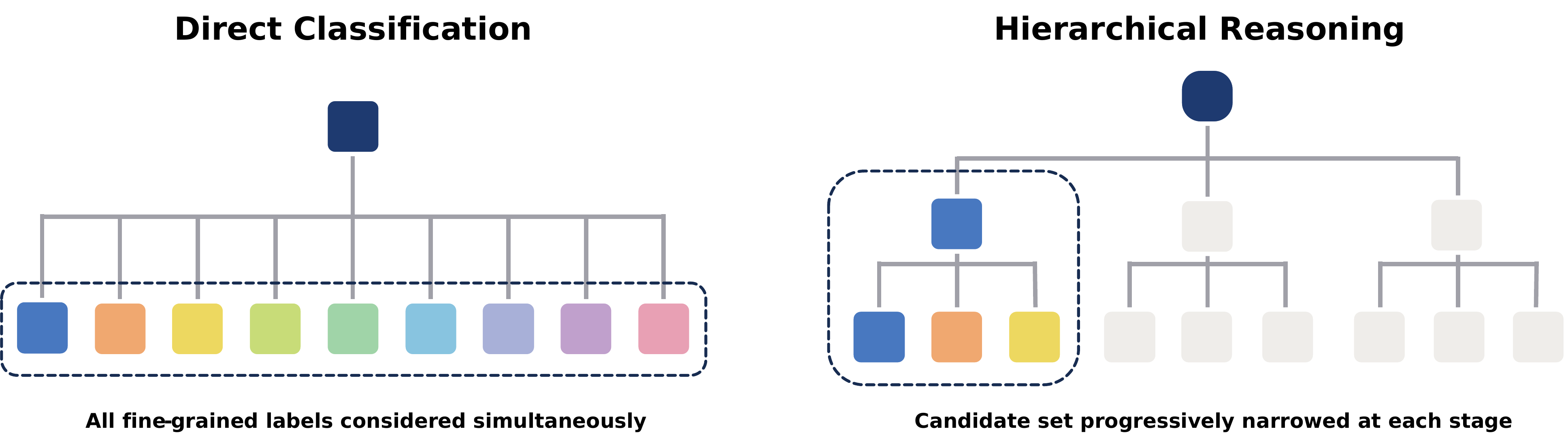}
    \caption{Comparison of Direct Classification and Hierarchical Reasoning}
    \label{fig.3}
\end{figure}

To address this challenge, we decompose the burden imposed by a large number of semantically related labels into a hierarchical structure. The detailed prompt is shown in Figure 2(C), when the framework is transferred to other actor-event triplets or other coding schemes, researchers need only replace the label definitions and disambiguation rules at each level to reuse the same hierarchical reasoning architecture. Rather than requiring the model to distinguish among all labels simultaneously, the hierarchical reasoning strategy progressively narrows the label set, allowing the model to focus only on a small number of proximate labels at each stage (see Figure \ref{fig.3}).

In our framework, with policy attitude as an example, we adopt the following procedure:

In the first step, the model identifies the first-level policy attitude through five diagnostic questions, each corresponding to one possible category: material cooperation, verbal cooperation, no policy or statement, verbal conflict, and material conflict. Concretely, the model is asked whether the text should be classified as each of these five categories in turn, for example, “should this text be classified as verbal cooperation?” or “should this text be classified as material conflict?” Each answer includes a brief rationale. These five question-answer pairs are then passed to an LLM formatter, which consolidates the reasoning outputs into a structured intermediate output. Rule-based Python code then uses this formatted first-level category to retrieve the corresponding second-level category space for the next step.

In the second step, the model assigns a more specific policy attitude label within the selected first-level category. For example, if the LLM formatter records the first-level category as material conflict, the next prompt presents is limited to relevant second-level categories, including downgrading or severing bilateral relations, hostile military actions, reneging on or downgrading treaties, restricting access, communication, and cooperation, and strengthening blockades or sanctions (the full list of second-level categories is reported in Online Appendix A). The model then chooses the most appropriate second-level label based on the category definitions. This narrower prompt makes the classification task more focused and helps the model compare only closely related options.

For label spaces of varying sizes, researchers can flexibly define the number of hierarchical levels as appropriate. However, excessively deep hierarchies may introduce cumulative error propagation across reasoning levels. The design of hierarchical label structures therefore requires balancing label-space reduction against the risk of cascading errors.

\section{Applying MDAF on Five Eyes Countries’ China Policy Dataset}
To demonstrate the applicability of the multi-dimensional annotation framework proposed in this paper to complex international relations analysis, we apply MDAF to the China policies of the Five Eyes countries: the U.S., the UK, Canada, Australia, and New Zealand.

This topic was selected not only because of its significance in contemporary international relations, but also because it reflects the multiple challenges of converting complex policy texts into structured data. Western countries' stances towards China across different policy domains frequently exhibit a complicated mix of cooperation and confrontation. The positions of a given country may also vary over time. This highly complex, cross-issue, and dynamically evolving policy dynamic is ideally suited to testing the multi-dimensional coding framework's capacity to handle intricate texts. 

\subsection{Dataset Construction}
\begin{figure}[htbp]
    \centering
    \includegraphics[width=1\textwidth]{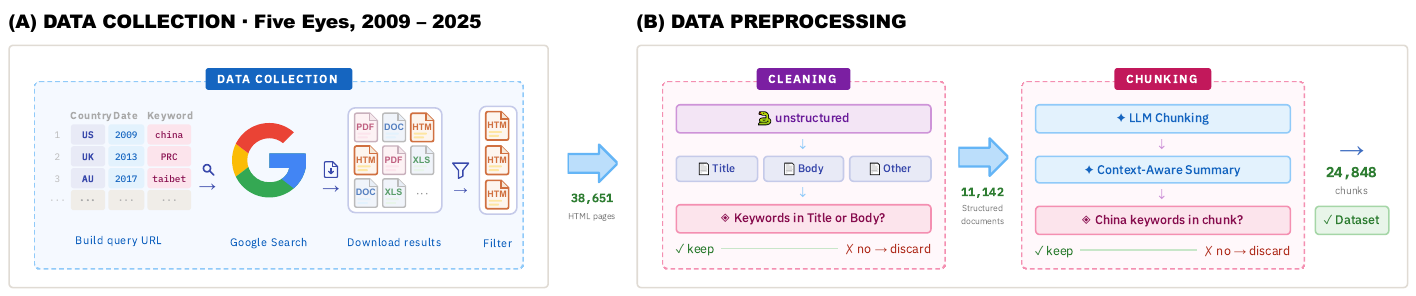}
    \caption{Pipeline of Dataset Construction}
    \label{fig.4}
\end{figure}

This section describes the construction of the dataset used in the empirical analysis. Figure~\ref{fig.4} summarizes the overall pipeline, from the systematic collection of China-related policy texts on official Five Eyes government websites to their preprocessing into annotation-ready text units.

\subsubsection{Collecting Raw Data via Google Advanced Search}
To construct a dataset of China-related policy texts from the Five Eyes countries, this study draws on official government websites as data sources and covers the period from January 2009 to February 2025.  Online queries were constructed along three dimensions (country, year, and China-related keywords\footnote{The China keyword list includes three types of terms: country-level identifiers such as “China,” “Chinese,” and “PRC”; political entities such as “CCP,” “Xi Jinping,” and “Hu Jintao”; and geographic terms associated with China-related policy discussions, such as “Beijing,” “Tibet,” “Xinjiang,” and “Hong Kong.” }) to generate structured query URLs, which were then used to search within the corresponding national government domains via Google Advanced Search.

Documents identified by the search procedure were downloaded in bulk and covered multiple formats, including PDF, DOC, HTML, and XLS. To ensure consistency in text extraction and downstream processing, we retained HTML pages as the raw corpus. PDF and DOC files often duplicated substantive textual content available in HTML form on the same official websites, while XLS files mainly contained tabular materials. This collection and filtering procedure yielded 38,651 HTML pages.

\subsubsection{Data Cleaning and Chunking}
The raw HTML pages needed to undergo two stages of preprocessing - cleaning and chunking - before they could be converted into text units suitable for input into the annotation pipeline.

For cleaning, we used the open-source package Unstructured to parse each HTML page into title, body, and other page elements. We then retained only pages in which China-related keywords appeared in the title or body. Pages where such keywords appeared only in navigation bars, footers, sidebars, or other non-content sections were discarded, as they did not contain substantive China-policy content. After cleaning, the dataset was reduced from 38,651 HTML pages to 11,142 valid pages.

Policy texts vary enormously in length, from brief press statements to strategic reports running to hundreds of pages, making it neither practical nor precise to treat entire documents as the unit of annotation. We then employed an LLM-based semantic chunking strategy, segmenting each document into a series of text portions according to thematic coherence, such that each chunk constitutes a relatively self-contained semantic unit. To preserve document-level contextual information that might otherwise be lost after chunking, we generated a context-aware summary for each chunk, compressing and attaching background information, such as the document title, source institution, and overarching theme, for use in subsequent annotation stages. The chunked output was then subjected to a further round of keyword filtering, retaining only those chunks containing China-related keywords in their text, and discarding passages within documents that were unrelated to China. Through chunking and filtering, a final output of 24,848 text chunks was produced, constituting the input dataset for the annotation framework. These chunks will subsequently processed to extract thematic actor-event triplets, which then underwent multi-dimensional annotation (as described in Sections 2 and 3 above).

\subsection{Evaluation}
\subsubsection{Human-Annotated Labels and Evaluation Baselines}
To evaluate annotation accuracy, we have constructed a human-annotated evaluation set through stratified sampling across country, document type, and chunk length, yielding 356 annotation units (see Online Appendix B for the full sampling protocol). Each unit was independently coded by three PhD candidates specialising in international politics, all of whom were trained on the codebook described in Section 2 prior to annotation. Inter-annotator agreement, measured by Krippendorff's $\alpha$, reached 0.907 on average. You can get detailed Krippendorff's $\alpha$ score per category in Online Appendix B. Where annotators disagreed, the majority label was adopted as the ground-truth; units on which all three annotators diverged were resolved through joint discussion. These consistently high agreement scores across all three dimensions reflect the precision and clarity of the coding framework: by providing exhaustive category definitions, mutually exclusive labels, and concrete annotation examples in the codebook, annotator discretion was substantially constrained, resulting in near-ceiling inter-rater reliability.

In Section 3, we have identified two problems: cross-dimensional interference between logically interrelated categorization schemes; and heterogeneity in the inference burden required by different annotation tasks. To systematically evaluate whether and to what extent MDAF addresses these problems, we have designed ablation studies, in which specific components of the framework are progressively removed to assess their contribution to overall performance. This design allows us to examine whether MDAF needs to combine all three strategies, rather than relying on only one or two of them. 

In addition, we introduce a single-prompt baseline, which is evaluated alongside MDAF on the same set of 356 manually annotated samples. The single-prompt baseline consolidates the annotation of all dimensions into a single LLM call, requiring the model to simultaneously produce complete annotations for all labels with a single prompt. This variant removes both the dimension-separated architecture and the reasoning steps, representing the most streamlined annotation scheme possible. It should be noted that the second-level categories of policy attitude were not included in the baseline evaluation: the sheer volume of label definitions at that level would have rendered the baseline model unable to produce valid structured output at all. Or, in other words, excluding sub-categorization ensures that the single-prompt baseline maintains a basic level of functional performance.

\subsubsection{Evaluation of the Need for Direct Classification}
To evaluate the need for direct classification (Section 3.1), Table~\ref{table1} compares annotation results for the state represented dimension using the decoupled reasoning strategy, the direct classification strategy, and the single prompt baseline. The results show that, even for very simple tasks, single prompting is suboptimal but direct classification is sufficient. Decoupled reasoning yields a marginal gain over direct classification of only 0.6\% in the F1 score. By applying direct classification to simple tasks of this kind, substantial resources can be saved without any significant sacrifice in accuracy. The example illustrates both the necessity and the effectiveness of incorporating the direct classification strategy into the MDAF.

\begin{table}[hbt!]
\begin{threeparttable}
\caption{Performance Comparison of Annotation Strategies for the State Represented Dimension}
\label{table1}
\begin{tabular}{llll}
\toprule
\headrow Strategy & ACC & F1 & MCC\\
\midrule
Decoupled Reasoning & 0.949 & 0.970 & 0.907\\
\midrule
Direct Classification & 0.951 & 0.964 & 0.906\\
\midrule
Single-Prompt Baseline & 0.885 & 0.885 & 0.770\\
\bottomrule
\end{tabular}
\begin{tablenotes}[hang]
\item[]\textit{Note}: ACC = accuracy; F1 = weighted F1 score; MCC = Matthews correlation coefficient.
\end{tablenotes}
\end{threeparttable}
\end{table}

\subsubsection{Evaluation of the Need for Decoupled Reasoning}
Table \ref{table2} presents the aggregated metrics for three primary classification dimensions: policy attitude, policy domain, and policy motivation. The results indicate that the decoupled reasoning strategy substantially improves precision on complex tasks compared to direct classification. The decoupled reasoning strategy significantly outperforms both direct classification and the single-prompt baseline across all three dimensions, with improvements generally ranging from 15 to 25 percentage points. Furthermore, direct classification and the single-prompt baseline exhibit severe underperformance on certain categories.

\begin{table}[hbt!]
\begin{threeparttable}
\caption{Performance Comparison of Annotation Strategies for the Policy Attitude, Policy Motivation, and Policy Domain Dimensions}
\label{table2}
\begin{tabular}{llccccc}
\toprule
\headrow Dimension & Variant & ACC & PREC & REC & F1\\
\midrule
Policy attitude
  & Decoupled Reasoning    & 0.956 & 0.887 & 0.897 & 0.891\\
  & Direct Classification  & 0.893 & 0.731 & 0.727 & 0.719\\
  & Single-Prompt Baseline & 0.879 & 0.718 & 0.618 & 0.657\\
\cmidrule(lr){1-6}
Policy motivation
  & Decoupled Reasoning    & 0.909 & 0.798 & 0.779 & 0.771\\
  & Direct Classification  & 0.896 & 0.714 & 0.527 & 0.594\\
  & Single-Prompt Baseline & 0.886 & 0.708 & 0.554 & 0.596\\
\cmidrule(lr){1-6}
Policy domain
  & Decoupled Reasoning    & 0.990 & 0.934 & 0.986 & 0.959\\
  & Direct Classification  & 0.956 & 0.853 & 0.783 & 0.804\\
  & Single-Prompt Baseline & 0.947 & 0.949 & 0.589 & 0.709\\
\bottomrule
\end{tabular}
\begin{tablenotes}[hang]
\item[]\textit{Note}: PREC = precision; REC = recall.
\end{tablenotes}
\end{threeparttable}
\end{table}

Table \ref{table3} presents the metrics for the three multi-class dimensions: stage, scope, and actor level. It can be seen that the decoupled reasoning strategy yields even greater improvements on multi-class metrics, with gains up to 32 percentage points. We attribute this larger improvement to the inherently higher decision complexity and category discrimination difficulty of multi-class tasks, which place greater demands on the model's reasoning capacity. LLMs struggle to concentrate their attention on distinguishing between multiple semantically similar categories with subtle boundaries. For instance, different policy stages often exhibit temporal and functional overlap, while different governance scopes present nested relationships. The reasoning-formatting strategy, by explicitly guiding the model to generate intermediate reasoning steps, enables it to progressively reduce the interference caused by semantic overlap between categories throughout the reasoning process.

\begin{table}[hbt!]
\begin{threeparttable}
\caption{Performance Comparison of Annotation Strategies for the Policy Stage, Policy Scope, and Actor Level Dimensions}
\label{table3}
\begin{tabular}{llccc}
\toprule
\headrow Dimension & Variant & ACC & F1 & MCC\\
\midrule
Policy stage
  & Decoupled Reasoning    & 0.930 & 0.934 & 0.901\\
  & Direct Classification  & 0.649 & 0.665 & 0.519\\
  & Single-Prompt Baseline & 0.609 & 0.639 & 0.488\\
\cmidrule(lr){1-5}
Policy scope
  & Decoupled Reasoning    & 0.876 & 0.881 & 0.804\\
  & Direct Classification  & 0.739 & 0.746 & 0.630\\
  & Single-Prompt Baseline & 0.744 & 0.763 & 0.604\\
\cmidrule(lr){1-5}
Actor level
  & Decoupled Reasoning    & 0.934 & 0.938 & 0.903\\
  & Direct Classification  & 0.725 & 0.706 & 0.556\\
  & Single-Prompt Baseline & 0.645 & 0.649 & 0.488\\
\bottomrule
\end{tabular}
\begin{tablenotes}[hang]
\item[]\textit{Note}: ACC = accuracy; F1 = weighted F1 score; MCC = Matthews correlation coefficient.
\end{tablenotes}
\end{threeparttable}
\end{table}

Table \ref{table4} provides a detailed performance of fine-grained category level performance under the decoupled reasoning strategy. It can be seen that F1 scores exceed 0.90 for the vast majority of policy domain categories, reflecting stable and well-balanced performance. For most categories, precision and recall are distributed evenly, yielding high F1 scores, demonstrating that decoupled reasoning strategy exhibits robustness and generalizability. These results indicate that the decoupled reasoning strategy not only improves overall performance, but also maintains stability when confronted with low-frequency and medium-inference-burden categories.

\begin{table}[hbt!]
\begin{threeparttable}
\caption{Fine-Grained Category-Level Performance under the Decoupled Reasoning Strategy}
\label{table4}
\begin{tabular}{llcccccr}
\toprule
\headrow Dimension & Label & ACC & PREC & REC & F1 & SUP & SUP\%\\
\midrule
Policy attitude
  & Material cooperation          & 0.961 & 0.867 & 0.897 & 0.881 & 58  & 16.30\\
  & Verbal cooperation            & 0.958 & 0.804 & 0.860 & 0.831 & 43  & 12.10\\
  & No policy or statement        & 0.966 & 0.806 & 0.853 & 0.829 & 34  & 9.60\\
  & Verbal conflict               & 0.924 & 0.905 & 0.933 & 0.918 & 163 & 45.80\\
  & Material conflict             & 0.969 & 0.963 & 0.852 & 0.904 & 61  & 17.10\\
\cmidrule(lr){1-8}
Policy motivation
  & Threat                        & 0.958 & 0.924 & 0.948 & 0.936 & 116 & 32.60\\
  & Economic interest             & 0.975 & 0.955 & 0.944 & 0.950 & 90  & 25.30\\
  & Alliance/alignment (source) & 0.624 & 0.201 & 0.738 & 0.316 & 42  & 11.80\\
  & Alliance/alignment (bilateral)& 0.941 & 0.855 & 0.783 & 0.817 & 60  & 16.90\\
  & Alliance/alignment (China)    & 0.823 & 0.424 & 0.241 & 0.308 & 58  & 16.30\\
  & Values ideology               & 0.941 & 0.980 & 0.835 & 0.901 & 115 & 32.30\\
  & Historical issue              & 0.992 & 1.000 & 0.250 & 0.400 & 4   & 1.10\\
  & Domestic politics             & 0.997 & 0.500 & 1.000 & 0.667 & 1   & 0.30\\
  & Order based                   & 0.868 & 0.500 & 0.532 & 0.515 & 47  & 13.20\\
  & Challenge                     & 0.975 & 0.936 & 0.880 & 0.907 & 50  & 14.00\\
\cmidrule(lr){1-8}
Policy domain
  & Trade economy and investment  & 0.969 & 0.918 & 0.968 & 0.942 & 93  & 26.10\\
  & Science technology and innovation & 0.989 & 0.902 & 1.000 & 0.949 & 37 & 10.40\\
  & Education                     & 0.994 & 0.889 & 1.000 & 0.941 & 16  & 4.50\\
  & Ideology                      & 0.986 & 0.955 & 1.000 & 0.977 & 105 & 29.50\\
  & Security and military         & 0.978 & 0.929 & 1.000 & 0.963 & 104 & 29.20\\
  & Climate change and environment& 0.997 & 0.960 & 1.000 & 0.980 & 24  & 6.70\\
  & Public health                 & 0.997 & 0.955 & 1.000 & 0.977 & 21  & 5.90\\
  & Sovereignty                   & 0.994 & 0.985 & 0.985 & 0.985 & 67  & 18.80\\
  & Cyber security                & 0.986 & 0.792 & 1.000 & 0.884 & 19  & 5.30\\
  & Law                           & 0.975 & 0.941 & 0.952 & 0.947 & 84  & 23.60\\
  & Travel migration and refugees & 0.994 & 0.933 & 0.933 & 0.933 & 15  & 4.20\\
  & Development aid and humanitarian & 0.997 & 0.857 & 1.000 & 0.923 & 6 & 1.70\\
  & Energy                        & 0.997 & 0.917 & 1.000 & 0.957 & 11  & 3.10\\
  & Infrastructure and transportation & 0.997 & 0.833 & 1.000 & 0.909 & 5 & 1.40\\
  & Culture media and sport       & 0.997 & 0.958 & 1.000 & 0.979 & 23  & 6.50\\
\bottomrule
\end{tabular}
\begin{tablenotes}[hang]
\item[]\textit{Note:} SUP = support count (number of labeled samples belonging to each category); SUP\% = support rate (the share of each category out of the total samples). 
The “Alliance/alignment reason” label is divided into three subtypes: Alliance/alignment (bilateral), where the source country’s stance toward China is shaped by its direct bilateral relationship with China; Alliance/alignment (source), where the source country’s stance is influenced by coordination with its own allies or partners, such as the UK aligning its China policy with the NATO, or the G7; and Alliance/alignment (China), where the source country responds to China’s coordination with a third party, such as the U.S. reacting to China’s cooperation with Russia. 
\end{tablenotes}
\end{threeparttable}
\end{table}

\subsubsection{Evaluation of Hierarchical Reasoning Strategy}
To evaluate the necessity and effectiveness of the hierarchical reasoning strategy (section 3.3), we compare its accuracy with that of decoupled reasoning (section 3.2) for the first and second level of policy attitude. Under decoupled reasoning, the two levels are wrapped into the same prompt whereas, under hierarchical reasoning, the prompt for the second level is conditional on the output for the first level. When calculating second-level performance, we restrict the evaluation to observations for which the first-level policy attitude is correctly classified, so that the second-level metrics capture errors in fine-grained categorisation rather than errors propagated from the first level.

As shown in Table  \ref{table5}, decoupled reasoning performs reasonably well for the first-level categories listed at the beginning of the prompt. This may be attributed to the format of the input classification rules, which provides more prominent cues for first-level labels, thereby attracting greater attention from the LLM. Another possible explanation is that first-level categories are semantically more straightforward and therefore easier for the model to select. Even here, hierarchical reasoning achieves higher accuracy (95.6\% vs. 89.3\%). For second-level categories appearing in the middle of the prompt, however, the accuracy of decoupled reasoning drops precipitously (from 89.3\% to 16.9\%). Here, hierarchical reasoning outperforms decoupled reasoning by a very large margin (76.6\% vs. 16.9\%). The same trends are observed for F1 scores and the MCC. The sharp difference between the first and second levels suggests that, in complex classification tasks, the LLM has limited capacity to allocate attention to information located in the middle of the prompt, making it difficult to cover a large number of categories in a stable manner. The stronger performance of the hierarchical reasoning strategy, by contrast, is likely due to the fact that large-label-space tasks is divided into distinct hierarchical levels. This optimizes the LLM’s attention allocation and simplifies the logical complexity at the second level.The hierarchical reasoning strategy keeps the overall error at a much lower level, resulting in higher accuracy, F1, and MCC scores. Although errors may propagate from one level into another, this appears manageable for up to three or maybe even four layers.

\begin{table}[hbt!]
\begin{threeparttable}
\caption{Performance Comparison of Hierarchical and Decoupled Reasoning Strategies for the Policy Attitude Dimension}
\label{table5}
\begin{tabular}{llccc}
\toprule
\headrow Level & Variant & ACC & F1 & MCC\\
\midrule
1st Level
  & Hierarchical Reasoning & 0.956 / 0.854 & 0.891 / 0.921 & 0.848 / 0.803\\
  & Decoupled Reasoning    & 0.893 / 0.730 & 0.719 / 0.844 & 0.641 / 0.627\\
\cmidrule(lr){1-5}
2nd Level
  & Hierarchical Reasoning & 0.766 & 0.792 & 0.642\\
  & Decoupled Reasoning    & 0.169 & 0.215 & 0.173\\
\cmidrule(lr){1-5}
Both Correct
  & Hierarchical Reasoning & 0.654 &   -   &   -  \\
  & Decoupled Reasoning    & 0.124 &   -   &   -  \\
\bottomrule
\end{tabular}
\begin{tablenotes}[hang]
\item[]\textit{Note:} 1st Level metrics are reported as 
per-label average / sample-level exact match.
\end{tablenotes}
\end{threeparttable}
\end{table}

\subsection{Findings}
This section presents selected preliminary findings from applying the proposed annotation scheme and automated annotation system to China-related policy texts from the Five Eyes countries over the study period. The analysis reveals fine-grained variation often obscured by aggregate accounts of China policy, providing a more nuanced view of how governments have adjusted their policies toward China.

\begin{figure}[htbp]
    \centering
    \includegraphics[width=1\textwidth]{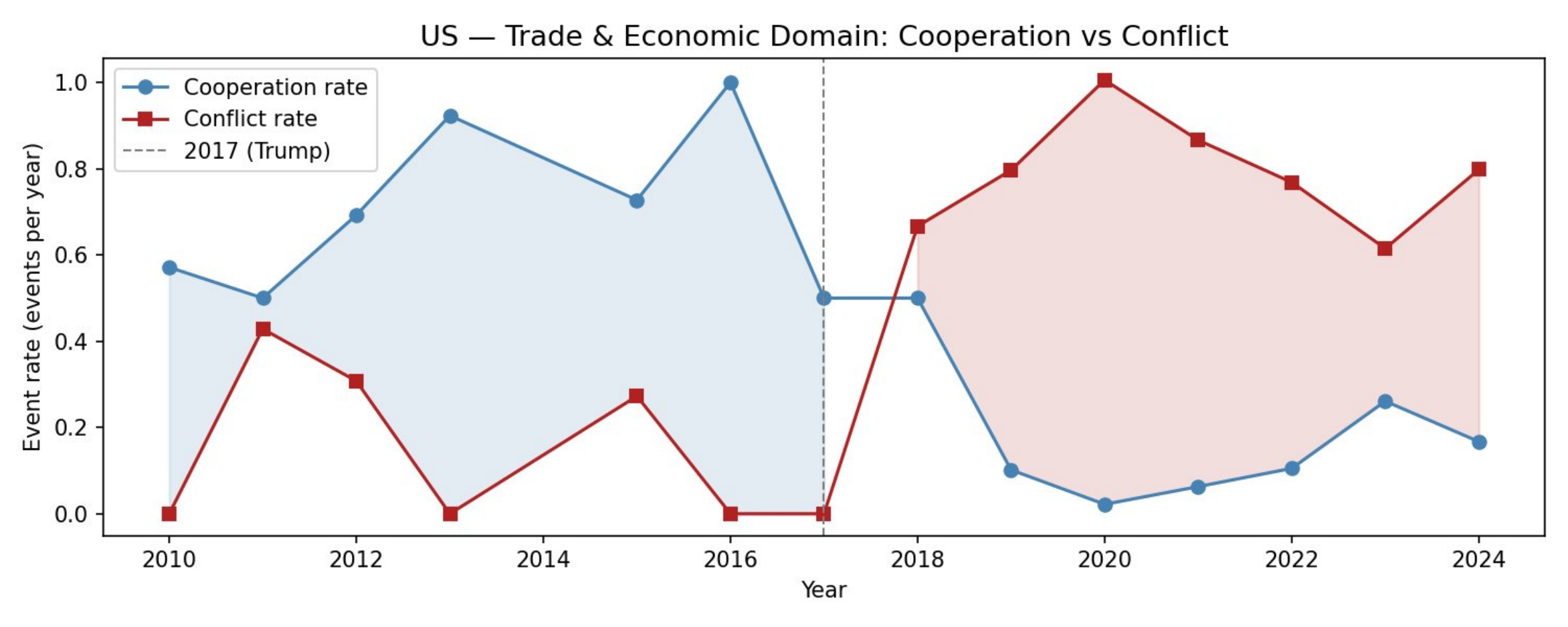}
    \caption{Cooperation and Conflict in U.S. Economic Policy towards China, 2010-2024}
    \label{fig.5}
\end{figure}

\begin{figure}[htbp]
    \centering
    \includegraphics[width=1\textwidth]{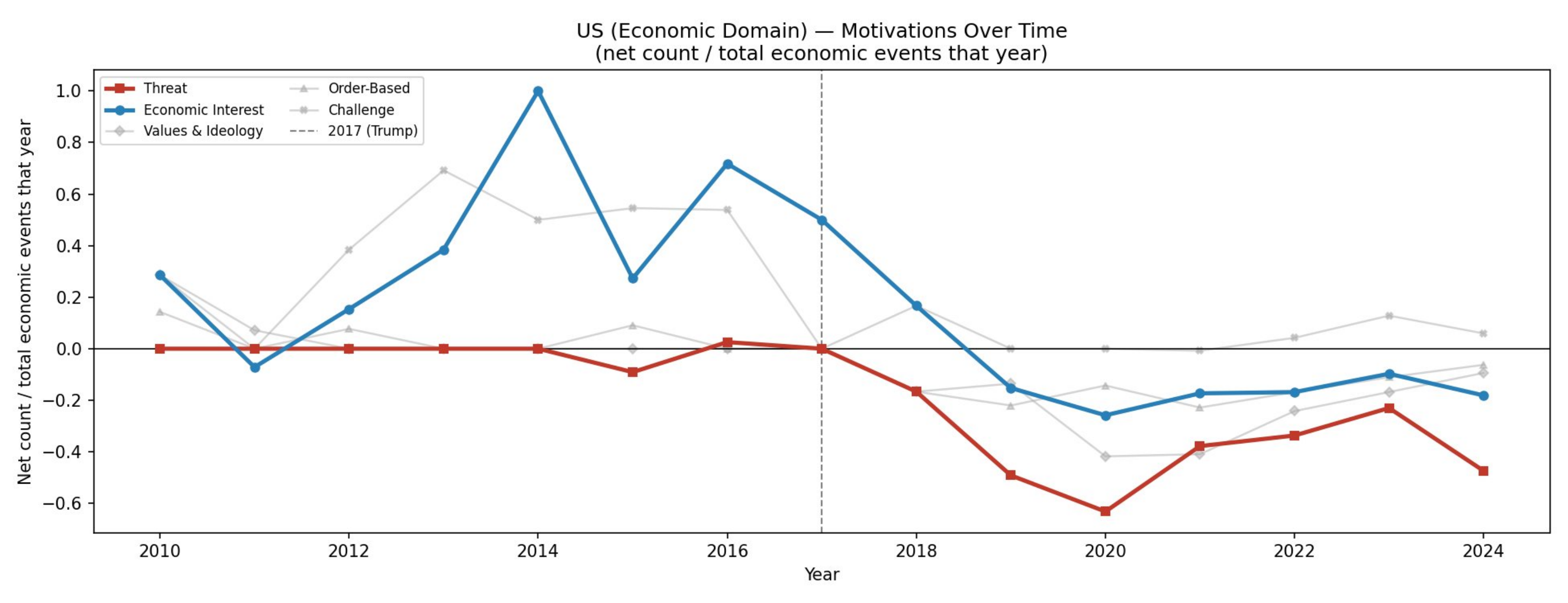}
    \caption{Motivational Shifts in U.S. Economic Policy towards China, 2010-2024}
    \label{fig.6}
\end{figure}

Figure \ref{fig.5} and Figure \ref{fig.6} use U.S. economic policy towards China to illustrate how the system links changes in policy orientation with shifts in underlying motivation. Figure \ref{fig.5} shows a clear reversal in policy attitude after 2017: cooperative signals decline sharply, while conflictual signals become increasingly dominant in the trade and economic domain. Figure \ref{fig.6} helps explain this shift by showing how the motivational basis of U.S. economic policy changed over time. Before 2017, economic interests were more often associated with engagement, reflecting concerns such as market access, commercial opportunity, and the benefits of economic interdependence. After 2017, however, economic issues were increasingly framed through the language of threat, competition, and vulnerability. Trade deficits, technology transfer, industrial competition, and supply-chain dependence were no longer treated only as economic problems, but became part of a broader economic security agenda. Taken together, the two figures show that the U.S. shift towards a more conflictual China policy was not simply a change in attitude. It also involved a reinterpretation of economic interests themselves, from sources of mutual benefit to sources of strategic risk.


\begin{figure}[htbp]
    \centering
    \includegraphics[width=1\textwidth]{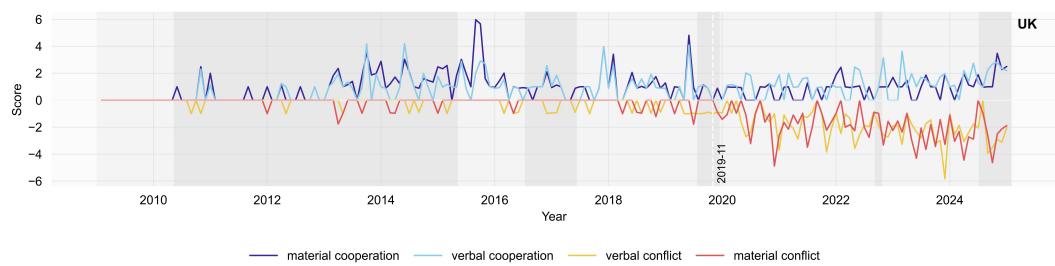}
    \caption{Aggregate Evolution of UK Policy Attitudes toward China, 2010-2024}
    \label{fig.7}
\end{figure}

\begin{figure}[htbp]
    \centering
    \includegraphics[width=1\textwidth]{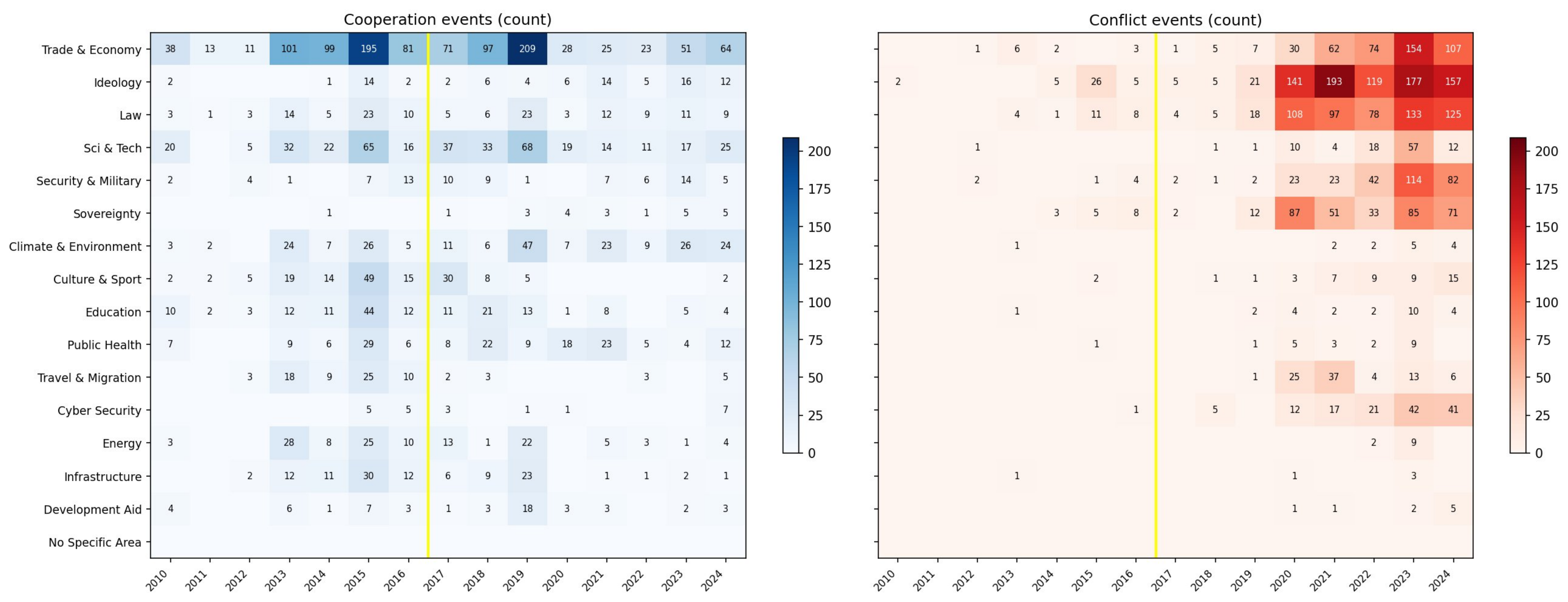}
    \caption{Domain-Level Distribution of UK Cooperation and Conflict toward China, 2010-2024}
    \label{fig.8}
\end{figure}

Figures \ref{fig.7} and \ref{fig.8} examine the evolution of the UK’s policy attitudes towards China between 2010 and 2024. Figure \ref{fig.7} shows that British China policy did not move in a single direction: positive and cooperative signals remained visible throughout the period, while negative and conflictual signals became more frequent and pronounced after 2019. Figure \ref{fig.8} maps colour codings for policy attitude upon policy domain labels. The results show that the balance of cooperation and conflict was not evenly distributed across UK China policies, but structured by issue areas. Domains such as climate \& environment, science \& technology, public health, and cultural \& sport continued to display predominantly positive signals. By contrast, negative signals became more concentrated in domains such as ideology, law, sovereignty, security \& military, cyber security, and travel \& immigration. The domain of trade and economy is particularly notable, not only because its framing shifted from earlier engagement-oriented language towards a more conflictual and risk-oriented pattern but also because, even as conflict signals increase, cooperate signals continue as a surprisingly high level.

\begin{figure}[htbp]
    \centering
    \includegraphics[width=1\textwidth]{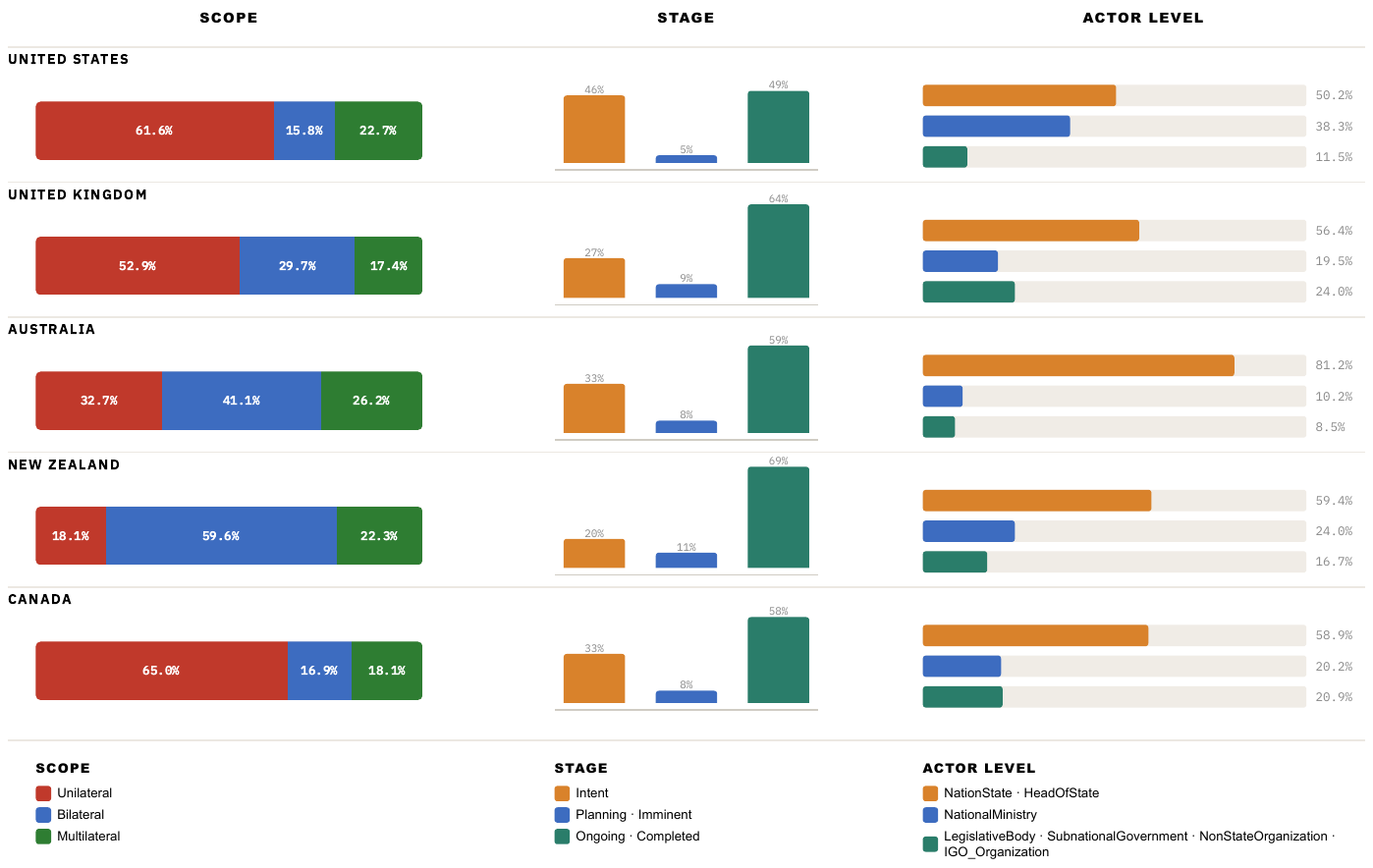}
    \caption{Policy Scope, Policy Stage, and Actor Level of China-Related Policy Events across the Five Eyes Countries, 2017-2024}
    \label{fig.9}
\end{figure}

Figure \ref{fig.9} compares China-related policy events across the Five Eyes countries from 2017 to 2024 along three structural dimensions: policy scope, policy stage, and actor level.

On the scope dimension, the five countries display distinct ways of framing China-related policy. Canada and the U.S. rely most heavily on unilateral framings (65.0\%; 61.6\%). The UK follows a similar but less pronounced pattern (52.9\%). By contrast, New Zealand stands out for its bilateral orientation, indicating greater emphasis on direct relationship management and issue-specific engagement with China (59.6\%). Australia is the most balanced case, combining unilateral, bilateral, and multilateral framing more evenly (32.7\%; 41.1\%; 26.2\%).

The policy-stage distribution indicates that most China-related texts refer to actions already in progress or completed. This pattern is most pronounced in New Zealand and the UK (69.1\%; 63.6\%), followed by Australia and Canada (59.0\%; 58.2\%). The U.S. is the main exception: although ongoing or completed actions remain the largest category (48.9\%), pure intent also accounts for a high share (45.9\%), suggesting a stronger presence of declaratory positioning and strategic signaling. Across all five countries, planning-related texts remain limited (5.2-10.6\%), indicating that public policy texts tend to emphasize either broad intentions or actions already under way, rather than intermediate stages of policy design or preparation.

The actor-level distribution captures who speaks and through which institutional channels China policy is expressed. Australia is the most centralized case: head-of-state and government-level actors account for the overwhelming majority of its texts (81.2\%), indicating that China policy is strongly concentrated at the level of national leadership and the state as a whole. The other four countries display less centralized patterns. In the U.S., ministerial actors play a prominent role in articulating China-related policy (38.3\%). New Zealand also has a relatively high ministerial share (24.0\%), consistent with its issue-management style of bilateral engagement. Canada and the UK show greater institutional plurality, with agency-level and subnational actors more visible in their China-policy texts (20.9\%; 24.0\%). This appears related to policy devolution from the central state to sub-state units such as Quebec in Canada and Scotland in the UK. 

\section{Limitations and Future Directions}
The framework constructed in this paper is built around the actor-event triplet. This triplet serves as the basic unit for multi-dimensional labeling and analysis. This design confers strong advantages when handling policy events with clear relational directionality, that is, cases in which an actor takes a policy action toward, or expresses a policy position on, a specific target. However, it also limits the framework’s adaptability to political processes or policy phenomena that are not easily reducible to actor-event triplet relations, such as the formation of policy preferences, internal negotiation over policy positions, or broader political dynamics that shape policy outputs without appearing as discrete relational events. Future research may explore ways to extend compatibility with non-relational policy phenomena while preserving the core structure of the current framework.

In addition, this study applies the same LLM across all annotation stages, regardless of task complexity. Yet the modular design of the workflow creates room for future model routing. Simpler tasks, such as information extraction, could be assigned to weaker but nimbler models, whereas tasks requiring more complex reasoning could be handled by stronger and weightier models. This strategy would help reduce computational costs while preserving annotation quality, thereby improving the scalability of the framework.

\section{Conclusion}
This article has developed MDAF, a multi-dimensional annotation framework for converting policy texts into structured policy event data. Our central argument is that automated policy annotation should not be treated as a single classification problem. Different annotation dimensions impose different inference burdens: direct classification is appropriate for low-inference-burden tasks, whereas medium-inference-burden tasks require decoupled reasoning and large-label-space tasks require hierarchical reasoning. MDAF addresses this variation by assigning different reasoning strategies within a single modular workflow. 

The evaluation section has demonstrated that this structured design substantially improves performance. Compared with direct classification, decoupled reasoning improves accuracy across key dimensions, including attitude, motivation, domain, scope, stage, and actor level, with gains ranging up to 32 percentage points. For large-label-space tasks, hierarchical reasoning is especially important, raising the accuracy of second-level attitude annotation from 16.9 percent to 76.6 percent. 

Our specific application of MDAF to China-related policy documents from the Five Eyes countries demonstrates the substantive value of the framework. By generating structured annotations across multiple dimensions, MDAF enables systematic comparison of how different governments define issues, express policy positions, justify decisions, and direct policy actions. It offers a scalable and fine-grained approach for the comparative study of policy discourse across countries and over time. 

MDAF contributes to international relations research and foreign policy analysis by preserving the richness of original texts while producing comparable and scalable data. More broadly, it shows that LLM-based annotation becomes more reliable when model reasoning is organized around the structure of the measurement task. Future work can extend the framework to non-foreign policy events and develop model-routing strategies to further improve reliability and scalability.

\paragraph{Competing Interests}
The authors declare none. 

\paragraph{Data Availability Statement}
The online appendix and replication code are available from the corresponding author upon reasonable request.

\printendnotes

\printbibliography

\end{document}